# High-Temperature Performance of MoS₂ Thin-Film Transistors: DC and Pulse Current-Voltage Characteristics

## C. Jiang[1,2], S. L. Rumyantsev[3,4], R. Samnakay[1,2] M.S. Shur[3] and A.A. Balandin[1,2,*]

[1]Nano-Device Laboratory (NDL), Department of Electrical Engineering, Bourns College of Engineering, University of California – Riverside, Riverside, California 92521 USA

[2]Phonon Optimized Engineered Materials (POEM) Center, Materials Science and Engineering Program, University of California – Riverside, Riverside, California 92521 USA

[3]Department of Electrical, Computer, and Systems Engineering, Center for Integrated Electronics, Rensselaer Polytechnic Institute, Troy, New York 12180, USA

[4]Ioffe Physical-Technical Institute, St. Petersburg 194021, Russia

### Abstract

The measurements of the high-temperature current-voltage characteristics of MoS₂ thin-film transistors show that the devices remain functional to temperatures of at least as high as 500 K. The temperature increase results in decreased threshold voltage and mobility. The comparison of the DC and pulse measurements shows that the DC sub-linear and super-linear output characteristics of MoS₂ thin-films devices result from the Joule heating and the interplay of the threshold voltage and mobility temperature dependences. At temperatures above 450 K, an intriguing phenomenon of the "memory step" – a kink in the drain current - occurs at zero gate voltage irrespective of the threshold voltage value. The memory step effect was attributed to the slow relaxation processes in thin films similar to those in graphene and electron glasses. The obtained results suggest new applications for MoS₂ thin-film transistors in extreme-temperature electronics and sensors.

*Corresponding author (AAB): balandin@ee.ucr.edu

## I. Introduction

Many electronic components for control systems and sensors are required to operate at temperature of ~500 K. Examples of the high temperature application include turbine engine control in aerospace and energy generation or the oil-field instrumentation. The availability of transistors and circuits for high temperature (T~500 K) operation is limited [1]. Devices made of SiC, GaN and other conventional large-band-gap semiconductors hold promise for extended high-temperature operation [2-4] but are still not cost-effective for high volume applications. There is a need for new material systems that can be used for fabrication of field-effect transistors (FETs), thin-film transistors (TFT) and chemical FET and TFT based sensors functional at high temperatures.

Recent advances in the exfoliation and growth of quasi two-dimensional (2D) layered materials – referred to as *van der Waals* materials – have resulted in a surge of interest to their possible device applications [5-8]. One of the most promising materials among the layered transition-metal dichalcogenides (TMD) is $MoS_2$ [9-10]. A single atomic layer of $MoS_2$ shows a direct band gap of $E_g$~1.9 eV while bi-layer and bulk $MoS_2$ exhibit an indirect band gap of ~1.6 eV and ~1.3 eV, respectively [11-13]. The interest to $MoS_2$ thin films and devices based on this material is explained by the relatively large energy band gap of $MoS_2$ [11-13], large on-off ratio of $MoS_2$ TFT [11-14], low levels of flicker noise in $MoS_2$ TFTs [15-17] and relatively high thermal conductivity of $MoS_2$ thin films compared to thin films of other TMDs [18]. It was also shown that $MoS_2$ TFT can be used as selective gas and chemical sensors [19-20].

In this paper, we report on the experimental demonstration of $MoS_2$ thin film transistors suitable for high-temperature applications. The $MoS_2$ TFT are functional to temperatures of *at least* as high as 500 K and remain operational for *at least* after a month of aging. Using the voltage pulse measurements, we were able to show that the sub-linear and super-linear output characteristics of $MoS_2$ TFT, frequently observed in the direct-current (DC) measurements, may result from the Joule heating and are associated with an interplay of the threshold voltage and mobility temperature dependences. We also report on the "memory step" effect – a kink in the drain current - occurring at zero gate voltage irrespective of the threshold voltage value. This effect observed at temperatures above 450 K can be used for $MoS_2$ thin film applications in high temperature sensors. The obtained results may lead to new applications for TFT implemented with $MoS_2$ and other similar van der Waals materials in extreme-temperature electronics and sensors.

## II. Material Characterization and Device Fabrication

For this study, thin films of $MoS_2$ were exfoliated from bulk crystals and transferred onto Si/SiO$_2$ substrates following the standard approach [5-8]. The thickness, H, of the exfoliated films ranged from a single layer to 18 layers. For the high-temperature experiments, we selected devices with the $MoS_2$ channel thickness $H= 9 − 12$ nm, which correspond to $15 − 18$ layers of $MoS_2$ (the thickness of a single layer of $MoS_2$ is 0.65 nm). The relatively thick films were more thermally stable and demonstrated a higher mobility at elevated temperatures. The gating was still possible, although it required an application of a higher back-gate voltage. The thickness and quality of thin films were determined with the atomic force microscopy (AFM) and micro-Raman spectroscopy. Figure 1 (a) shows a representative scanning electron microscopy (SEM) image of the studied $MoS_2$ TFT.

Micro-Raman spectroscopy (Renishaw InVia) was performed in the backscattering configuration under $\lambda$=488-nm laser excitation laser using an optical microscope (Leica) with a 100× objective. The excitation laser power was limited to less than 0.5 mW to avoid local heating. In Figure 1 (b), we present the informative bands at 383 cm$^{-1}$ ($E^1_{2g}$) and 406 cm$^{-1}$ ($A_{1g}$), consistent with the previous reports of the $MoS_2$ Raman spectra [21]. The Raman spectrum confirms that this sample is multilayer $MoS_2$ film. The latter follows from the frequency difference, $\Delta\omega$, between the $E^1_{2g}$ and the $A_{1g}$ peaks. The increase in the number of layers in $MoS_2$ films is accompanied by the red shift of the $E^1_{2g}$ and blue shift of the $A_{1g}$ peaks [21]. The inset shows the AFM image of a typical device.

[Figure 1: SEM, AFM and Raman]

Devices with $MoS_2$ channels were fabricated using electron beam lithography (LEO SUPRA 55) for patterning of the source and drain electrodes and the electron-beam evaporation (Temescal BJD-1800) for metal deposition. Conventional Si substrates with 300-nm thick $SiO_2$ layers were spin coated (Headway SCE) and baked consecutively with two positive resists: first, methyl methacrylate (MMA) and then, polymethyl methacrylate (PMMA). These devices consisted of $MoS_2$ thin-film channels with Au (100-nm) contacts. The heavily doped $Si/SiO_2$ wafer served as a back gate. The majority of the devices had a channel length, $L$, in the range from 1.3 μm to 3.5 μm, and the channel width, $W$, in the range from 1 μm to 7 μm.

### III. DC Characteristics of $MoS_2$ TFTs

We investigated direct-current (DC) characteristics of a number of $MoS_2$ TFT with the channel thickness of 2-3 atomic layers and 15 − 18 atomic layers. The devices were also tested after a few days and two month aging under ambient conditions. Figure 2 shows a comparison of the room temperature (RT) current-voltage (I-V) characteristics of the $MoS_2$ TFTs with the 2-3-layer and 15-18-layer thick channels. The effect of aging is also indicated. One can see that the $MoS_2$ TFTs with thicker channels are characterized by a higher threshold voltage, $V_{th}$, higher electron mobility, $\mu$, and smaller sub-threshold slope.

[Figure 2: Transfer Characteristics at RT]

Figure 3 presents transfer the characteristics of the $MoS_2$ TFTs for the temperature varying from RT to 500 K. The I-Vs were measured at the drain-source voltage $V_d$=0.1 V. The temperature increase leads to the shift of the threshold voltage to more negative values and decreasing mobility. A similar trend was found for the measurements at temperatures below RT [22]. An intriguing feature observed in the high-temperature I-V characteristics is a characteristic kink in the transfer current-voltage characteristics at T>450K. It was well reproduced at several different temperatures above T=450 K and during multiple measurements. This kink is seen at $V_g$=0 for different temperatures and, therefore, for different threshold voltages. A similar effect, which we called a "memory step" was found earlier for graphene transistors at elevated temperatures [23]. In graphene, the "memory step" was negative, i.e. current decreased step-like when scanning the gate voltage from negative to positive values, while in $MoS_2$ TFTs the I-Vs revealed a positive "memory step". A similar memory effect (called the "memory dip") was observed in electronic glasses [24-27]. Although the nature of this phenomenon is still a subject of debates, we can point out two common features in all three material systems: the onset of the effect at $V_g$=0 and presence of the slow relaxation processes [23-24]. We discuss this effect in more details below.

[Figure 3: Temperature Dependence of I-Vs]

The effective mobility, $\mu$, was calculated as [28]

$$\mu = \frac{L_g}{C_{OX} R_{ds} (V_g - V_{th}) W} \ .$$
(1)

Here $C_{OX} = \varepsilon_o \varepsilon_r / d = 1.15 \times 10^{-4}$ (F/m$^2$) is the oxide capacitance, where $\varepsilon_o$ is the dielectric permittivity of free space, $\varepsilon_r$ is the dielectric constant and $d$ is the oxide thickness. We used $\varepsilon_r$=3.9 and $d$=300 nm for the SiO$_2$ layer. Plotting the drain-to-source resistance, $R_{ds}$, vs. $1/(V_g-V_{th})$, and extrapolating this dependence to zero yielded the estimate for the total contact resistance. The extracted data indicated that the contact resistance was negligible for all the samples at all temperatures. Figure 4 presents the evolution of the effective mobility and the threshold voltage with temperature. The threshold voltage, $V_{th}$, only weakly depends on temperature for 300 K < T < 370 K. For T > 370 K, the mobility decreases rapidly indicating a high interface trap density.

[Figure 4: Mobility and $V_{th}$ Dependence on T]

In the linear regime, the drain-source current, $I_d$, is proportional to $\mu C_{OX}(V_g - V_{th})V_d$. The dependences of the mobility and the threshold voltage on temperature have an opposite effect on current − voltage characteristics: the decrease of the mobility leads to the current decrease but decrease of threshold voltage leads to the current increase. As a result, current can either increase or decrease with the temperature increase depending on the specific shape of mobility and threshold voltage temperature dependences. For this reason, the sub-linear and super-linear output characteristics, often observed in DC measurements with $MoS_2$ TFTs at drain voltage exceeding 1-2 V for the micron-size $MoS_2$ devices, can be attributed to Joule heating of the channel during the measurements and to the dominance of one or the other mechanism. Figure 5 shows the gate voltage dependence of the mobility at different temperatures. The weak gate voltage dependence of the mobility calculated assuming zero contact resistance is a piece of evidence suggesting a low contact resistance in the examined devices.

In this set of measurements, we established that $MoS_2$ TFTs are functional to temperature at least as high as 500 K. Since this is the first study of $MoS_2$ TFT performance at temperatures around 500 K, no direct comparison with other works [22, 29-31] is possible. However, there have been data reported for $MoS_2$ TFTs for temperatures around 350 K. In our devices (see Figure 4), the mobility decreased from 55 $cm^2$/Vs to 11 $cm^2$/Vs as the temperature increases from RT to 500 K. The same trend was observed for CVD $MoS_2$ devices characterized at temperatures between 150 K and 350 K [22].

[Figure 5: Mobility vs. Gate Voltage]

Practical applications of $MoS_2$ TFT in control circuits or sensors require that they operate at least longer than one month. Figure 6 shows the temperature dependences of the threshold voltage and mobility for $MoS_2$ TFTs aged over a month under ambient conditions. The aged devices were characterized by a higher threshold voltage, lower mobility and weaker temperature dependence of the mobility. As we pointed out already, the temperature dependence of the threshold voltage and mobility have an opposite effect on current voltage. In order verify this conclusion we performed pulse measurements. Sending microsecond pulses through $MoS_2$ TFTs allows one to avoid Joule heating.

[Figure 6: DC Characteristics of Aged Devices]

## IV. Pulse Characteristics of MoS$_2$ TFTs

In Figure 7 (a), we show the output current-voltage characteristics of $MoS_2$ TFT measured in the pulse mode at T=300 K and $V_g$=76 V. Figures 7 (b-c) present the shape of the 10-μs voltage and current pulses. The maximum electric field achieved in these tests was 86.5 kV/cm with the maximum current density for 10-nm thick channel of $3\times10^5$ A/cm$^2$. One can see that the current-voltage characteristic is still linear at these voltage and current levels. The current rise time at 300 K is ~0.5 μs. A slight increase of the current on the top of the pulse, which is about 0.5%, is attributed to the minor heating of the device channel.

[Figure 7: Pulse Characteristics at RT]

In order to observe the saturation in output current-voltage $MoS_2$ TFT characteristics the pulse measurements were performed at elevated temperature when the threshold voltage shifts to smaller values. Figure 8 (a) shows the output current-voltage characteristics measured in the pulse mode at the temperature T=453 K and $V_g$=0. Figure 8 (b) presents the shapes of the voltage and the current pulses at $V_g$=76 V. At 453 K the characteristic current rise time is ~5 μs. The current saturation is clearly seen in Figure 8 (a).

[Figure 8: Pulse Characteristics at Elevated Temperature]

## V. "Memory Step" Effects in $MoS_2$ TFTs

We now turn to the discussion of a possible origin of the "memory steps", which we observed in this work in $MoS_2$ TFTs and, previously, in graphene devices [23]. The exact origin of the observed memory step effect in $MoS_2$ devices with different $V_{th}$ and the reasons for its intriguing appearance near $V_g$=0 are not clear. At the same time, it is difficult not to notice a similarity between the observed feature in $MoS_2$ thin films with those in graphene [23] and electron glasses [24-27]. The electron glasses demonstrated memory effects including the so-called "memory dips", i.e., irregularity, referred to as the "dip", in the transfer I-V characteristics. Several mechanisms responsible for the "memory dip" have been discussed [24-27]. However, the underlying physics of the "memory dip" in the electronic glass is still under debates. In three different material systems – $MoS_2$ thin films, graphene and electron glasses – the current anomaly is found at $V_g$=0. However in contrast to $MoS_2$ thin films and graphene, the memory dips in electron glasses have been observed at cryogenic temperatures. In $MoS_2$ thin films and graphene the effect is found at elevated temperatures. All three materials are characterized by slow relaxation processes. In graphene and $MoS_2$ these are processes of capture and release of free carries by deep levels. The concentration of these deep levels can be high and comparable with the concentration of free careers. We believe that when characteristic times of capture and release are similar and comparable with the sweep time the condition for memory step is satisfied.

The specific step observed in Figure 3 at $V_g$=0 V can be explained by different mechanisms. The first proposed model is based on the following two assumptions: (i) the concentration of defects (deep traps) is comparable with the concentration of free carriers; and (ii) the characteristic times of the charge carrier trapping by defects, de-trapping and voltage sweep are of the same order of magnitude. In this case, when $MoS_2$ TFT stays for a long time at $V_g$=0 V, there will be some number of free carriers and a large number of trapped carriers. During the voltage sweep, a negative gate voltage is applied and then swiped towards positive values. During this sweep, the concentration of free carriers is lower than the equilibrium owing to the long de-trapping time. This results in the lower than steady state current. Since gate voltage reached some vicinity around $V_g$=0 the concentration of traps with electrons is below the equilibrium, which is compensated with the electrons from the contacts and step occurs. Given that the capture time is comparable with the voltage sweep time, the excess electrons ensure larger current above $V_g$ till certain voltage (about $V_g$=15 V in Figure 3).

The second model is based on the assumption of different trapping and de-trapping times. At low temperatures (including RT for this material system), the trapping and de-trapping times are very large. Therefore, the trap occupancy corresponds to that at the zero gate bias because the samples have been stored for long period of times (hours and days) with no gate bias applied. At all other biases, the steady-state trap occupancy is not achieved during the relatively fast sweep (minutes). At very high temperatures (around 450 K for given $MoS_2$ films), the de-trapping times become faster but trapping times remain about the same. Hence, when the Fermi level corresponds to the zero bias at high temperature, the traps can release additional charge resulting in the "memory step." Additional experiments and qualitative models are required to distinguish, which mechanism describes the observed phenomenon more accurately.

## VI. Conclusions

We investigated the high-temperature current-voltage characteristics of $MoS_2$ TFTs using DC and pulse measurements. The $MoS_2$ TFTs remained functional to temperatures of, at least, as high as 500 K. The temperature increase resulted in decreased threshold voltage and mobility, which have an opposite effect on the current-voltage characteristics. We established that the sub-linear and super-linear output characteristics of $MoS_2$ thin-films devices, frequently observed in DC measurements, may result from the Joule heating and interplay of the threshold voltage and mobility temperature dependences. The pulse measurements performed for the electric field of up to ~87-kV/cm at RT showed that the current-voltage characteristics are still linear. The current saturation in pulse measurements was achieved at $V_{SD}$~6 V when the device temperature was elevated to 453 K. At temperatures above 450 K, we observed an intriguing phenomenon of the "memory step" – a kink in the drain current that occurs at zero gate voltage. The memory step effect was attributed to the slow relaxation processes in thin films of $MoS_2$ similar to those in graphene and electron glasses.

*Acknowledgements*

This work was supported, in part, by the Semiconductor Research Corporation (SRC) and Defense Advanced Research Project Agency (DARPA) through STARnet Center for Function Accelerated nanoMaterial Engineering (FAME). AAB also acknowledges funding from the National Science Foundation (NSF) for the project Graphene Circuits for Analog, Mixed-Signal, and RF Applications (NSF CCF-1217382). SLR acknowledges partial support from the Russian Fund for Basic Research (RFBR). The work at RPI was supported by NSF under the auspices of the EAGER program. The authors thank Professor Y. Galperin, Professor V. Kachorovskii, and Professor M. Levinshtein, for helpful discussions.

**FIGURE CAPTIONS**

**Figure 1:** Atomic force microscopy image (a) and a line scan (b) of a representative $MoS_2$ TFT. Raman spectrum of $MoS_2$ thin film showing the $E^1_{2g}$ and the $A_{1g}$ peaks (c). The increase in the number of layers in $MoS_2$ films is accompanied by the red shift of the $E^1_{2g}$ and blue shift of the $A_{1g}$ peaks. The energy difference between $E^1_{2g}$ and the $A_{1g}$ peaks indicates that the given sample is a few-layer $MoS_2$ film.

**Figure 2:** Current-voltage characteristics of the fabricated $MoS_2$ FET with different channel thickness at room temperature. Note that TFTs with thicker $MoS_2$ channels are characterized by the higher threshold voltage, higher mobility and smaller sub-threshold slope. The transfer current-voltage characteristics for the few-days and two-month aged devices are also shown for comparison.

**Figure 3:** Temperature dependence of the transfer current-voltage characteristics at the drain-source voltage $V_d$=0.1 V. The temperature increase leads to the shift of the threshold voltage to more negative values. Note a kink in the drain current – referred to as "memory step" – reproducibly appearing at $V_g$=0 V for temperatures above T=450 K. The intriguing phenomenon was previously observed for electronic glasses and graphene.

**Figure 4:** Temperature dependences of the threshold voltage and mobility in the as fabricated $MoS_2$ TFT. The mobility and threshold temperature dependences have the opposite effect on the current-voltage characteristics.

**Figure 5:** Gate-voltage dependence of the effective mobility in $MoS_2$ TFT at different temperatures. The weak temperature dependence of the mobility suggests a low contact resistance.

**Figure 6:** Temperature dependences of the threshold voltage (a) and mobility (b) in the one-month aged $MoS_2$ TFT.

**Figure 7:** Output current-voltage characteristics of $MoS_2$ TFT measured in the pulse mode for $V_g$=76 V at room temperature (a). The shapes of 10-μs voltage and current pulses are shown in the panels (b) and (c), respectively.

**Figure 8:** Output current-voltage characteristics of $MoS_2$ TFT measured in the pulse mode for $V_g$=0 K at the elevated temperature of T=453 K (a). The shapes of the voltage and current pulses at $V_g$=76 V is shown in the panel (b). The characteristic current rise time at 435 K is ~5 μs.

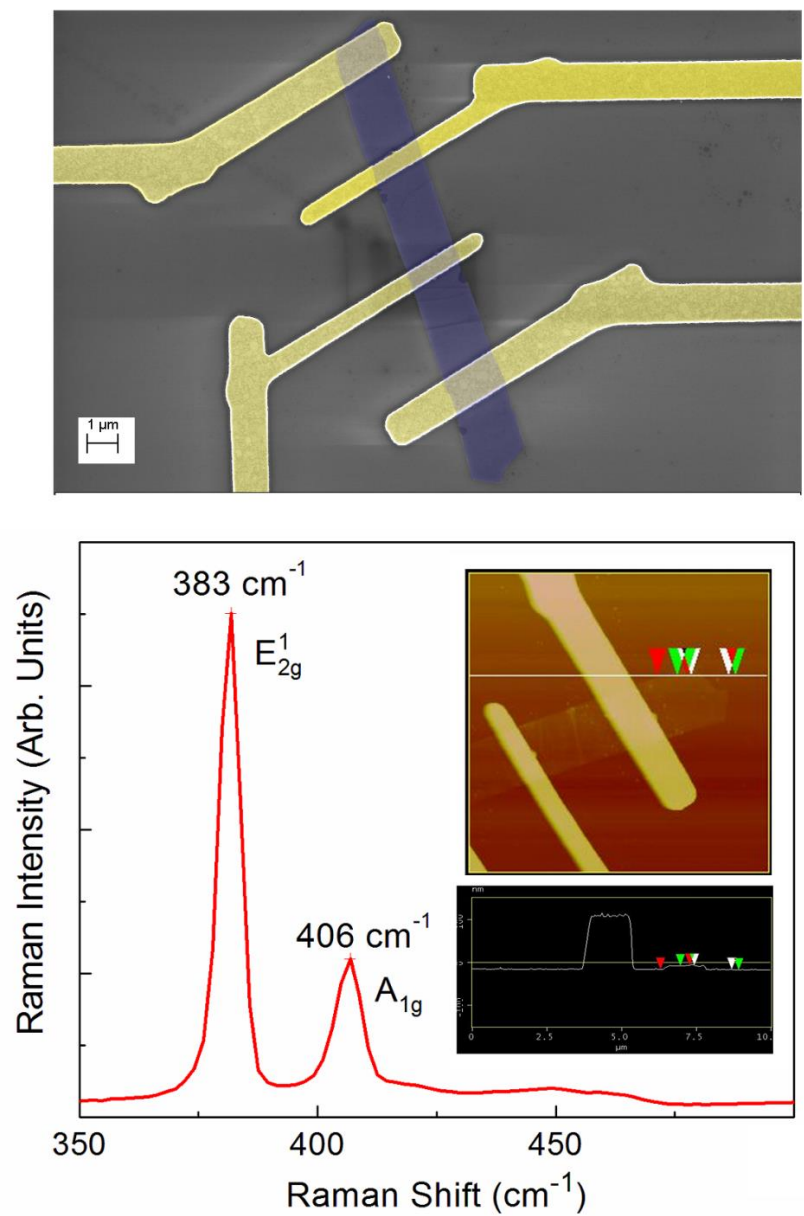

Figure 1

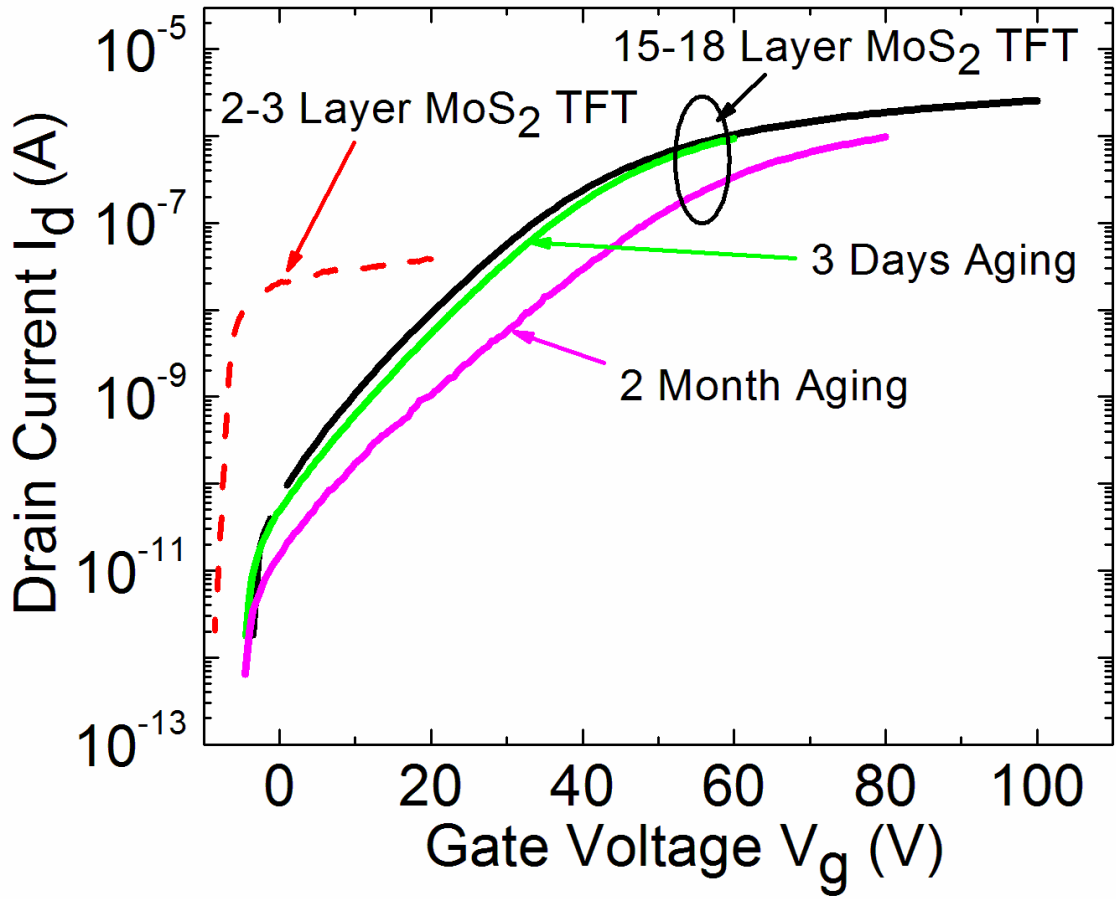

Figure 2

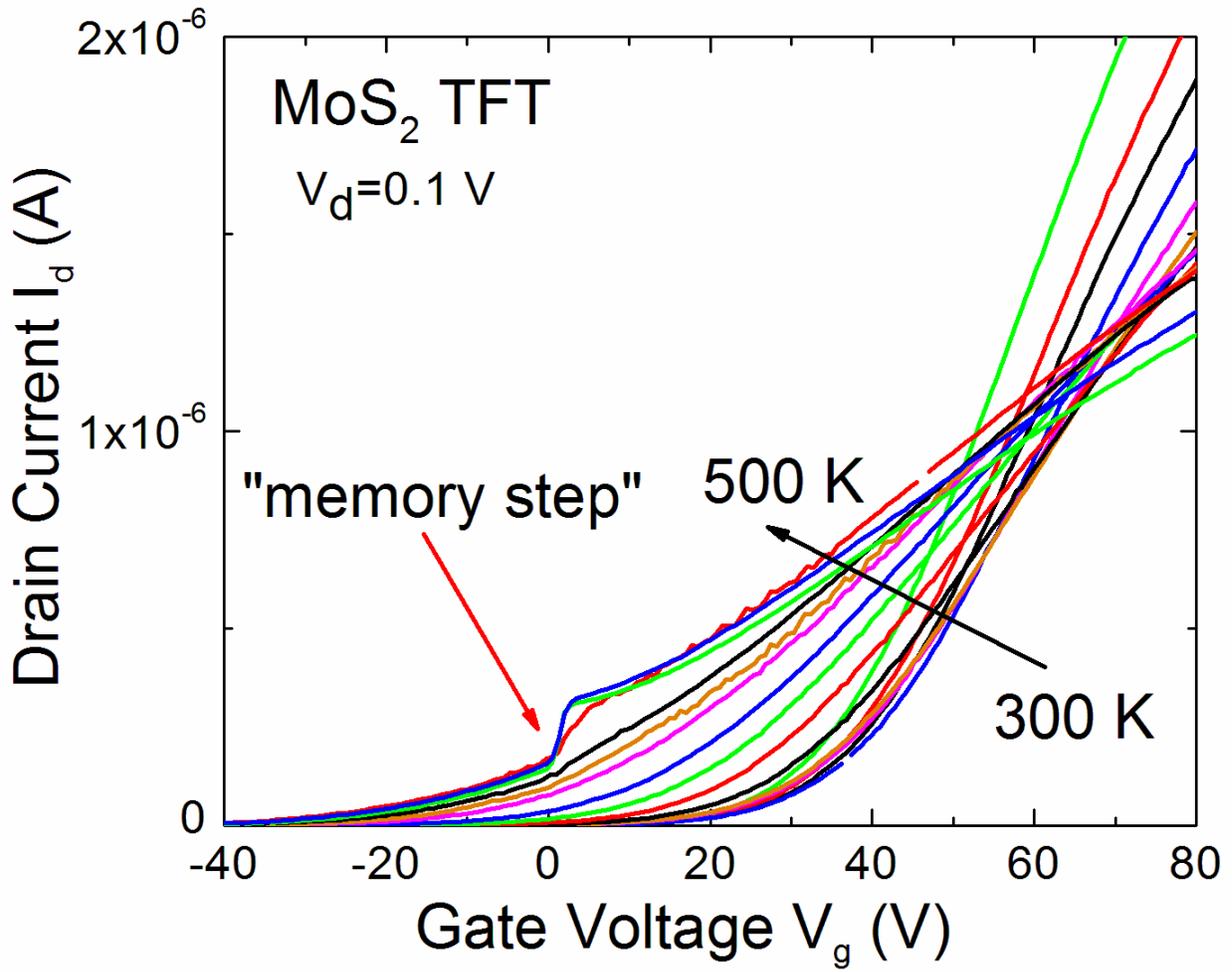

Figure 3

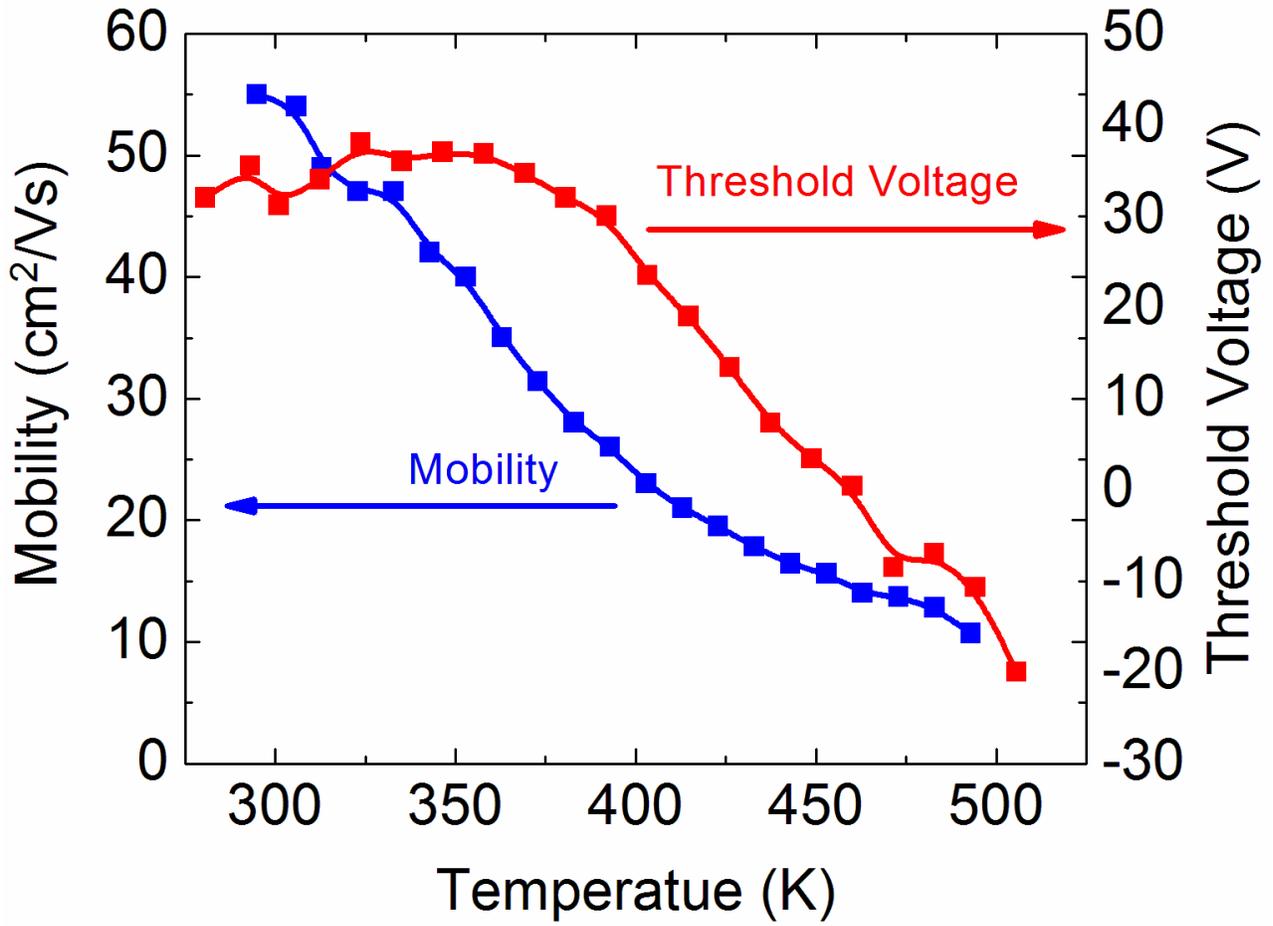

Figure 4

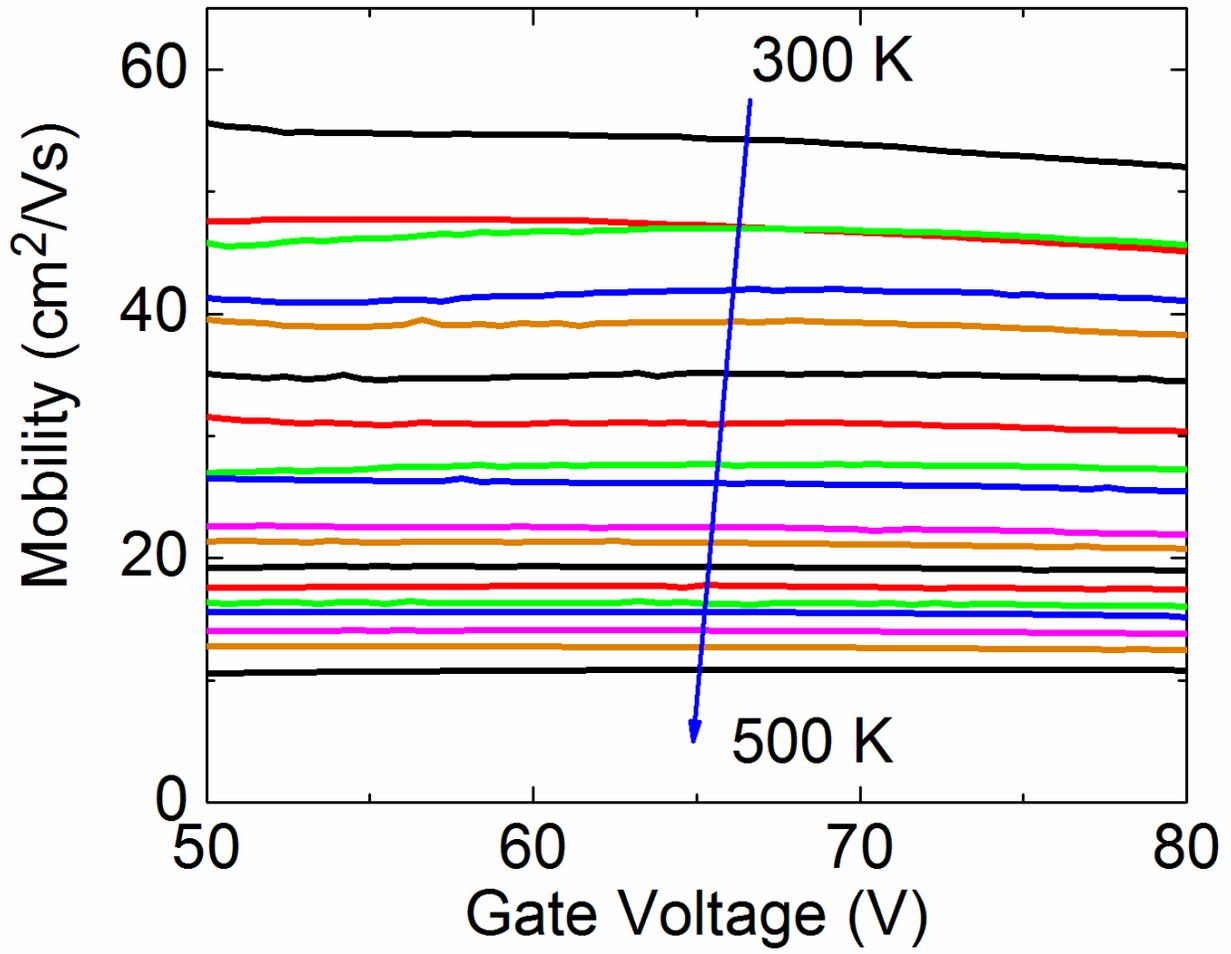

Figure 5

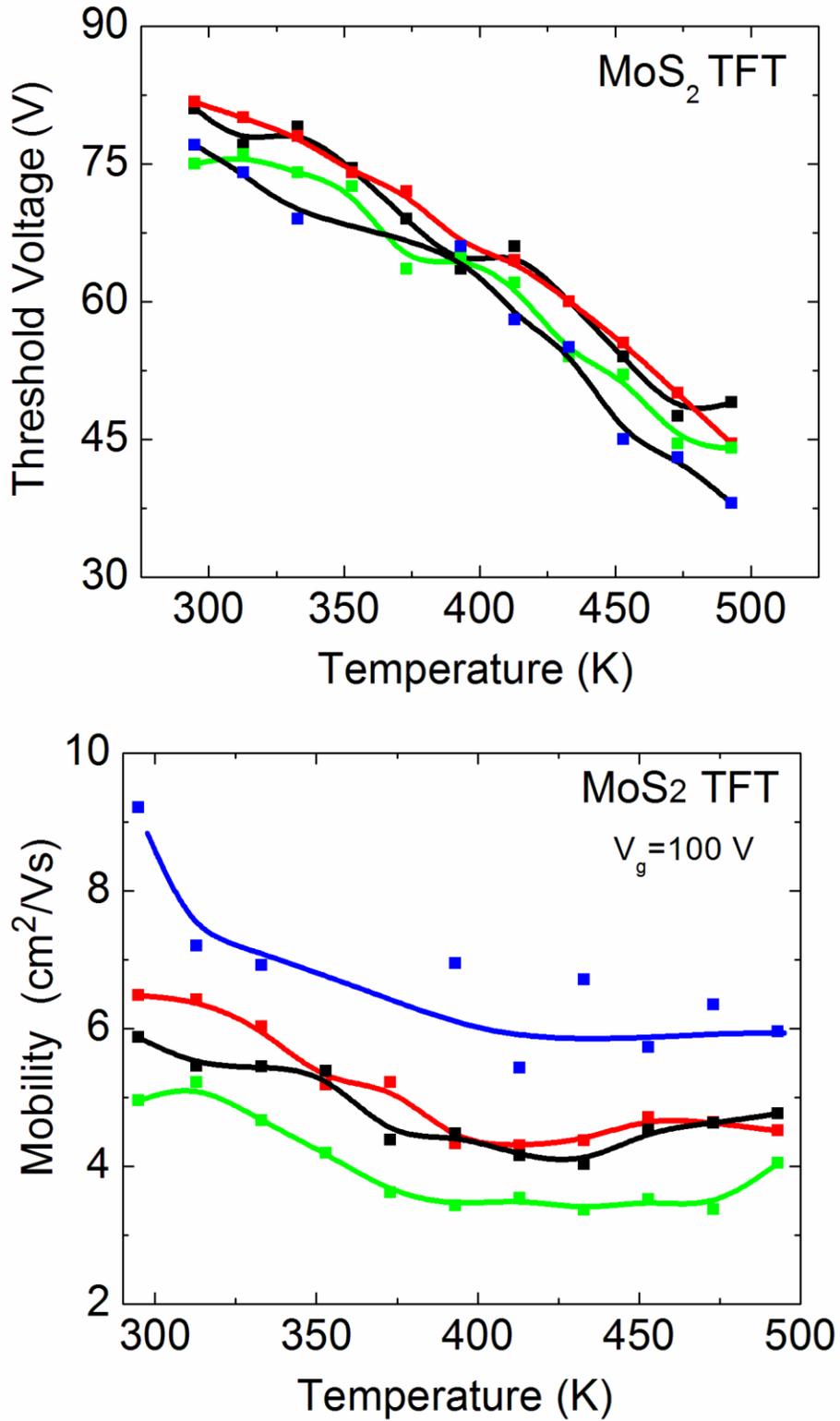

Figure 6

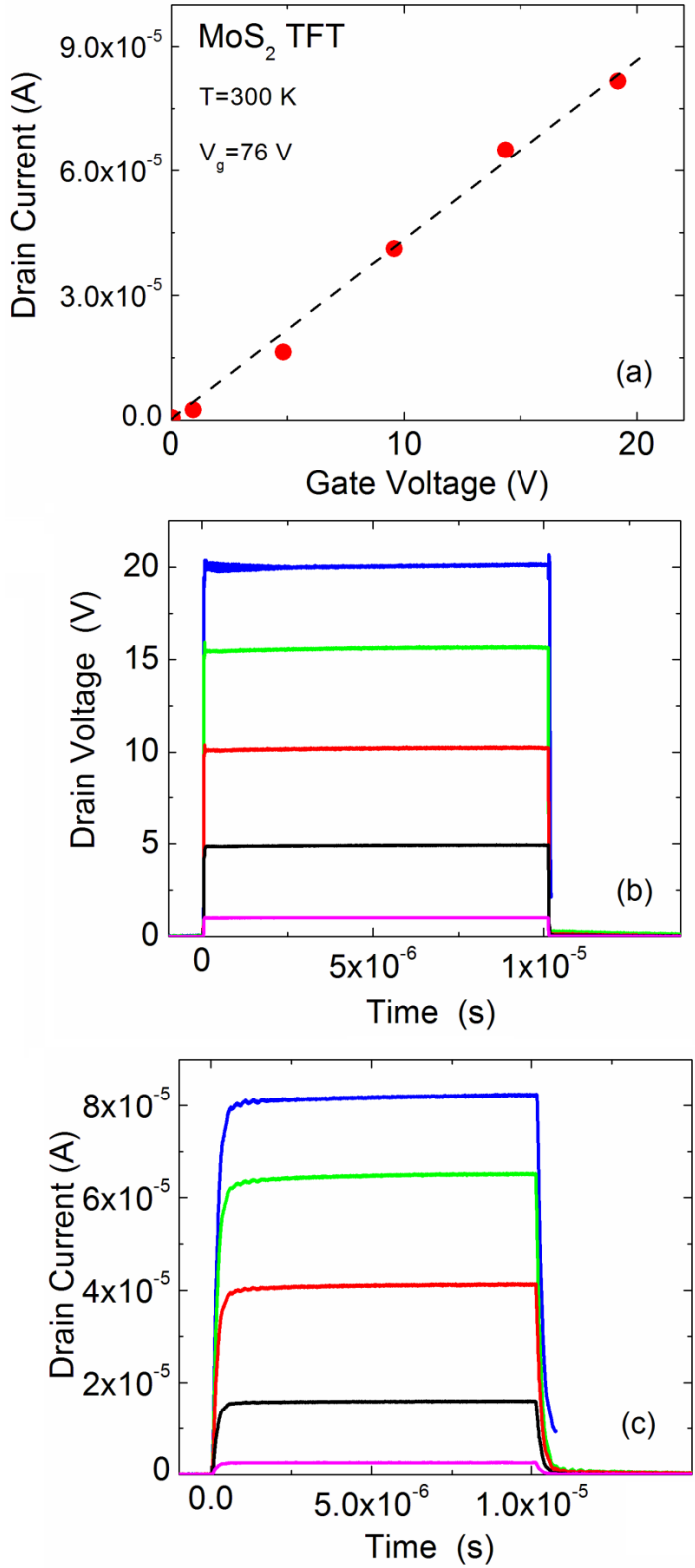

Figure 7

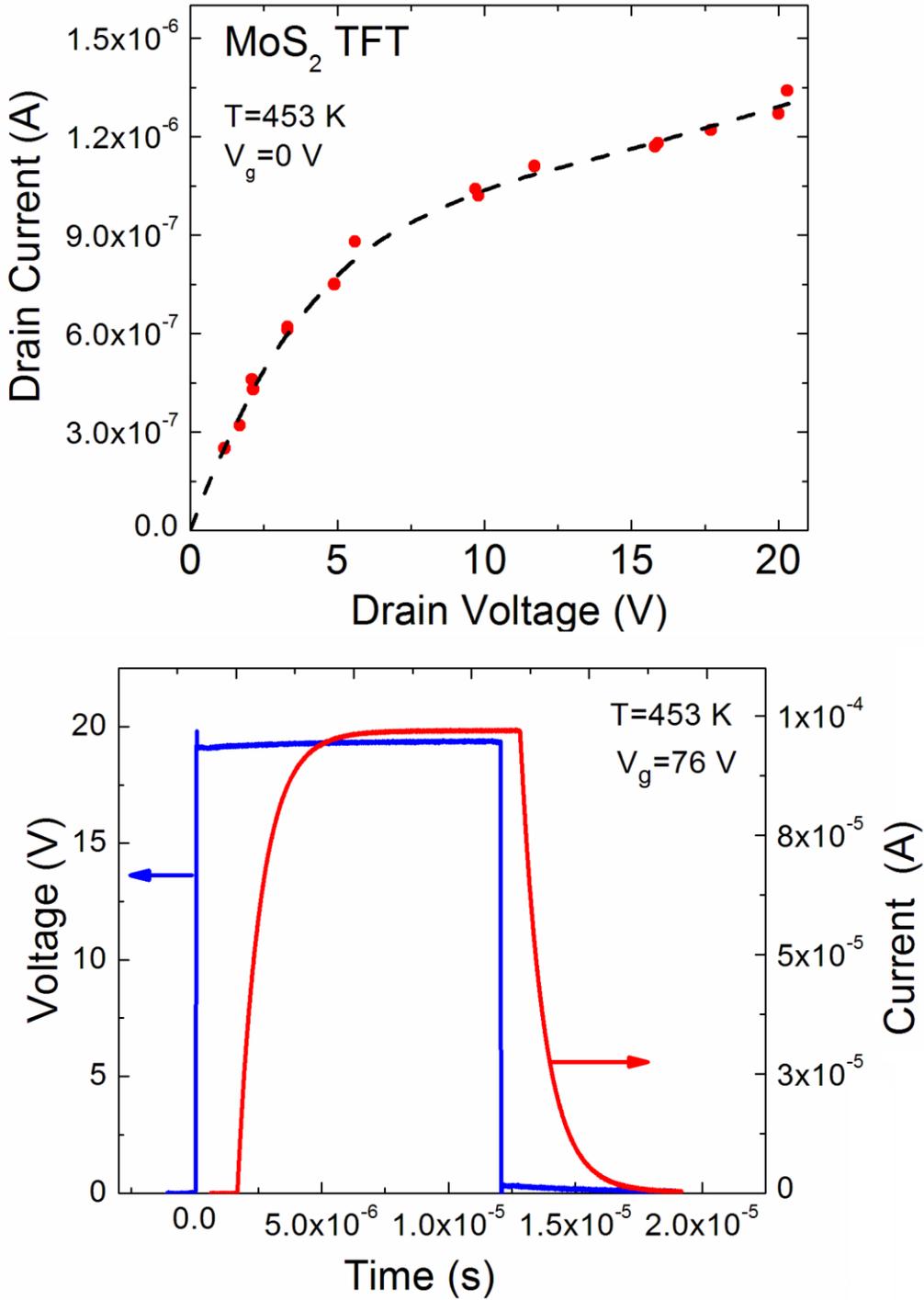

Figure 8